\documentclass{webofc}
\usepackage[varg]{txfonts}   
\begin{document}
\title{Galactic Star Formation with NIKA2 (GASTON):}
\subtitle{Evidence of mass accretion onto dense clumps}

\author{\firstname{A.~J.}~\lastname{Rigby}\inst{\ref{Cardiff}}\fnsep\thanks{\email{RigbyA@cardiff.ac.uk}}
  \and \firstname{R.}~\lastname{Adam} \inst{\ref{LLR}}
  \and  \firstname{P.}~\lastname{Ade} \inst{\ref{Cardiff}}
  \and  \firstname{H.}~\lastname{Ajeddig} \inst{\ref{CEA}}
  \and  \firstname{M.}~\lastname{Anderson} \inst{\ref{Cardiff}}
  \and  \firstname{P.}~\lastname{Andr\'e} \inst{\ref{CEA}}
  \and \firstname{E.}~\lastname{Artis} \inst{\ref{LPSC}}
  \and  \firstname{H.}~\lastname{Aussel} \inst{\ref{CEA}}
  \and  \firstname{A.}~\lastname{Bacmann} \inst{\ref{IPAG}}
  \and  \firstname{A.}~\lastname{Beelen} \inst{\ref{IAS}}
  \and  \firstname{A.}~\lastname{Beno\^it} \inst{\ref{Neel}}
  \and  \firstname{S.}~\lastname{Berta} \inst{\ref{IRAMF}}
  \and  \firstname{L.}~\lastname{Bing} \inst{\ref{LAM}}
  \and  \firstname{O.}~\lastname{Bourrion} \inst{\ref{LPSC}}
  \and  \firstname{A.}~\lastname{Bracco} \inst{\ref{Zagreb}}
  \and  \firstname{M.}~\lastname{Calvo} \inst{\ref{Neel}}
  \and  \firstname{A.}~\lastname{Catalano} \inst{\ref{LPSC}}
  \and  \firstname{M.}~\lastname{De~Petris} \inst{\ref{Roma}}
  \and  \firstname{F.-X.}~\lastname{D\'esert} \inst{\ref{IPAG}}
  \and  \firstname{S.}~\lastname{Doyle} \inst{\ref{Cardiff}}
  \and  \firstname{E.~F.~C.}~\lastname{Driessen} \inst{\ref{IRAMF}}
  \and  \firstname{P.}~\lastname{Garc\'ia} \inst{\ref{China}, \ref{Chile}}
  \and  \firstname{A.}~\lastname{Gomez} \inst{\ref{CAB}}
  \and  \firstname{J.}~\lastname{Goupy} \inst{\ref{Neel}}
  \and  \firstname{F.}~\lastname{K\'eruzor\'e} \inst{\ref{LPSC}}
  \and  \firstname{C.}~\lastname{Kramer} \inst{\ref{IRAME}}
  \and  \firstname{B.}~\lastname{Ladjelate} \inst{\ref{IRAME}}
  \and  \firstname{G.}~\lastname{Lagache} \inst{\ref{LAM}}
  \and  \firstname{S.}~\lastname{Leclercq} \inst{\ref{IRAMF}}
  \and  \firstname{J.-F.}~\lastname{Lestrade} \inst{\ref{LERMA}}
  \and  \firstname{J.-F.}~\lastname{Mac\'ias-P\'erez} \inst{\ref{LPSC}}
  \and  \firstname{A.}~\lastname{Maury} \inst{\ref{CEA}}
  \and  \firstname{P.}~\lastname{Mauskopf} \inst{\ref{Cardiff},\ref{Arizona}}
  \and  \firstname{F.}~\lastname{Mayet} \inst{\ref{LPSC}}
  \and  \firstname{A.}~\lastname{Monfardini} \inst{\ref{Neel}}
  \and  \firstname{M.}~\lastname{Mu\~noz-Echeverr\'ia} \inst{\ref{LPSC}}
  \and  \firstname{N.}~\lastname{Peretto}\inst{\ref{Cardiff}}
  \and  \firstname{L.}~\lastname{Perotto} \inst{\ref{LPSC}}
  \and  \firstname{G.}~\lastname{Pisano} \inst{\ref{Cardiff}}
  \and  \firstname{N.}~\lastname{Ponthieu} \inst{\ref{IPAG}}
  \and  \firstname{V.}~\lastname{Rev\'eret} \inst{\ref{CEA}}
  \and  \firstname{I.}~\lastname{Ristorcelli} \inst{\ref{Toulouse}}
  \and  \firstname{A.}~\lastname{Ritacco} \inst{\ref{IAS}, \ref{ENS}}
  \and  \firstname{C.}~\lastname{Romero} \inst{\ref{Pennsylvanie}}
  \and  \firstname{H.}~\lastname{Roussel} \inst{\ref{IAP}}
  \and  \firstname{F.}~\lastname{Ruppin} \inst{\ref{MIT}}
  \and  \firstname{K.}~\lastname{Schuster} \inst{\ref{IRAMF}}
  \and  \firstname{S.}~\lastname{Shu} \inst{\ref{Caltech}}
  \and  \firstname{A.}~\lastname{Sievers} \inst{\ref{IRAME}}
  \and  \firstname{C.}~\lastname{Tucker} \inst{\ref{Cardiff}}
  \and  \firstname{E.~J.}~\lastname{Watkins} \inst{\ref{Heidelberg}}
  \and  \firstname{R.}~\lastname{Zylka} \inst{\ref{IRAMF}}
}

\institute{
  School of Physics and Astronomy, Cardiff University, Queen’s Buildings, The Parade, Cardiff, CF24 3AA, UK 
  \label{Cardiff}
  \and
  LLR (Laboratoire Leprince-Ringuet), CNRS, \'{E}cole Polytechnique, Institut Polytechnique de Paris, Palaiseau, France
  \label{LLR}
  \and
  AIM, CEA, CNRS, Universit\'e Paris-Saclay, Universit\'e Paris Diderot, Sorbonne Paris Cit\'e, 91191 Gif-sur-Yvette, France
  \label{CEA}
  \and
  Univ. Grenoble Alpes, CNRS, IPAG, 38000 Grenoble, France 
  \label{IPAG}
  \and
  Univ. Grenoble Alpes, CNRS, Grenoble INP, LPSC-IN2P3, 53, avenue des Martyrs, 38000 Grenoble, France
  \label{LPSC}
  \and
  Institut d'Astrophysique Spatiale (IAS), CNRS, Universit\'e Paris Sud, Orsay, France
  \label{IAS}
  \and
  Institut N\'eel, CNRS, Universit\'e Grenoble Alpes, France
  \label{Neel}
  \and
  Institut de RadioAstronomie Millim\'etrique (IRAM), Grenoble, France
  \label{IRAMF}
  \and
  Aix Marseille Univ, CNRS, CNES, LAM (Laboratoire d'Astrophysique de Marseille), Marseille, France
  \label{LAM}
  \and 
  Rudjer Bo\v{s}kovi\'{c} Institute, Bijeni\v{c}ka cesta 54, 10000 Zagreb, Croatia
  \label{Zagreb}
  \and 
  Dipartimento di Fisica, Sapienza Universit\`a di Roma, Piazzale Aldo Moro 5, I-00185 Roma, Italy
  \label{Roma}
  \and 
  Chinese Academy of Sciences South America Center for Astronomy, National Astronomical Observatories, CAS, Beijing 100101, PR China
  \label{China}
  \and
  Instituto de Astronom\'ia, Universidad Cat\'olica del Norte, Av. Angamos 0610, Antofagasta 1270709, Chile
  \label{Chile}
  \and
  Centro de Astrobiolog\'ia (CSIC-INTA), Torrej\'on de Ardoz, 28850 Madrid, Spain
  \label{CAB}
  \and  
  Instituto de Radioastronom\'ia Milim\'etrica (IRAM), Granada, Spain
  \label{IRAME}
  \and 
  LERMA, Observatoire de Paris, PSL Research University, CNRS, Sorbonne Universit\'e, UPMC, 75014 Paris, France  
  \label{LERMA}
  \and
  School of Earth and Space Exploration and Department of Physics, Arizona State University, Tempe, AZ 85287, USA
  \label{Arizona}
  \and
  Univ. Toulouse, CNRS, IRAP, 9 Av. du Colonel Roche, BP 44346, 31028, Toulouse, France
  \label{Toulouse}
  \and 
  Laboratoire de Physique de l’\'Ecole Normale Sup\'erieure, ENS, PSL Research University, CNRS, Sorbonne Universit\'e, Universit\'e de Paris, 75005 Paris, France 
  \label{ENS}
  \and
  Department of Physics and Astronomy, University of Pennsylvania, 209 South 33rd Street, Philadelphia, PA, 19104, USA
  \label{Pennsylvanie}
  \and 
  Institut d'Astrophysique de Paris, Sarbonne Universit\'{e}, CNRS (UMR7095), 75014 Paris, France
  \label{IAP}
  \and 
  Kavli Institute for Astrophysics and Space Research, Massachusetts Institute of Technology, Cambridge, MA 02139, USA
  \label{MIT}
  \and
  Caltech, Pasadena, CA 91125, USA
  \label{Caltech}
  \and 
  Astronomisches Rechen-Institut, Zentrum f\"{u}r Astronomie der Universit at Heidelberg, Monchhofstraße 12-14, D-69120 Heidelberg, Germany
  \label{Heidelberg}
}

\abstract{%
  High-mass stars ($m_* \gtrsim 8 \, M_\odot$) play a crucial role in the evolution of galaxies, and so it is imperative that we understand how they are formed. We have used the New IRAM KIDs Array 2 (NIKA2) camera on the Institut de Radio Astronomie Millim\'{e}trique (IRAM) 30-m telescope to conduct high-sensitivity continuum mapping of $\sim2$ deg$^2$ of the Galactic plane (GP) as part of the Galactic Star Formation with NIKA2 (GASTON) large program. We have identified a total of 1467 clumps within our deep 1.15 mm continuum maps and, by using overlapping continuum, molecular line, and maser parallax data, we have determined their distances and physical properties. By placing them upon an approximate evolutionary sequence based upon 8 $\mu$m \textit{Spitzer} imaging, we find evidence that the most massive dense clumps accrete material from their surrounding environment during their early evolution, before dispersing as star formation advances, supporting clump-fed models of high-mass star formation.
}

\maketitle

\section{Introduction}
\vspace{-1mm}
\label{intro}
Obtaining an understanding of the origin of the stellar initial mass function (IMF) is one of the most important goals of contemporary astrophysics. Over the last decade, the advent of \textit{Herschel} observations have highlighted the link between dense filamentary structures within the interstellar medium (ISM) and star formation \citep{Molinari2010, Andre2010}. Thermally supercritical filaments are susceptible to gravitational fragmentation along their primary axis, resulting in strings of dense cores \citep[e.g.][]{Inutsuka1997} that may explain the shape of the IMF in the mass range $0.1 \lesssim m_* \lesssim 5\,\mathrm{M}_\odot$ \citep{Lee2017, Andre2019}. However, this \textit{core-fed} scenario may be insufficient outside these mass limits; some observations \citep[e.g.][]{Peretto2013, Chen2019} and simulations \citep[e.g.][]{Wang2010, Vazquez-Semadeni2019} suggest that high-mass star formation requires the large infall rates found at the centres of globally-collapsing molecular clouds in a \textit{clump-fed} scenario, while brown dwarf-formation may require processes such as dynamical ejection from multiple star systems \citep{Bate2002} or circumstellar disc fragmentation \citep{Thies2010}.

To tackle the origin of the IMF at the low- and high-mass extremes, the 200-hour GASTON large program (PI: N. Peretto) is conducting deep mapping of a region of the Galactic plane that is rich in high-mass star formation, in addition to surveying more nearby clouds for pre-brown dwarfs, whilst exploiting the dual waveband capabilities of NIKA2 on the IRAM 30-m telescope to study potential environmental variations in dust properties. In this proceeding, we summarise recent results from the GP survey \citep{Rigby2021}, focusing on the transition between the core-fed and clump-fed modes of star formation.

\section{Observations}
\vspace{-1mm}
\label{obs}

\begin{figure*}
\vspace{-7mm}
\begin{center}
\includegraphics[width=\textwidth]{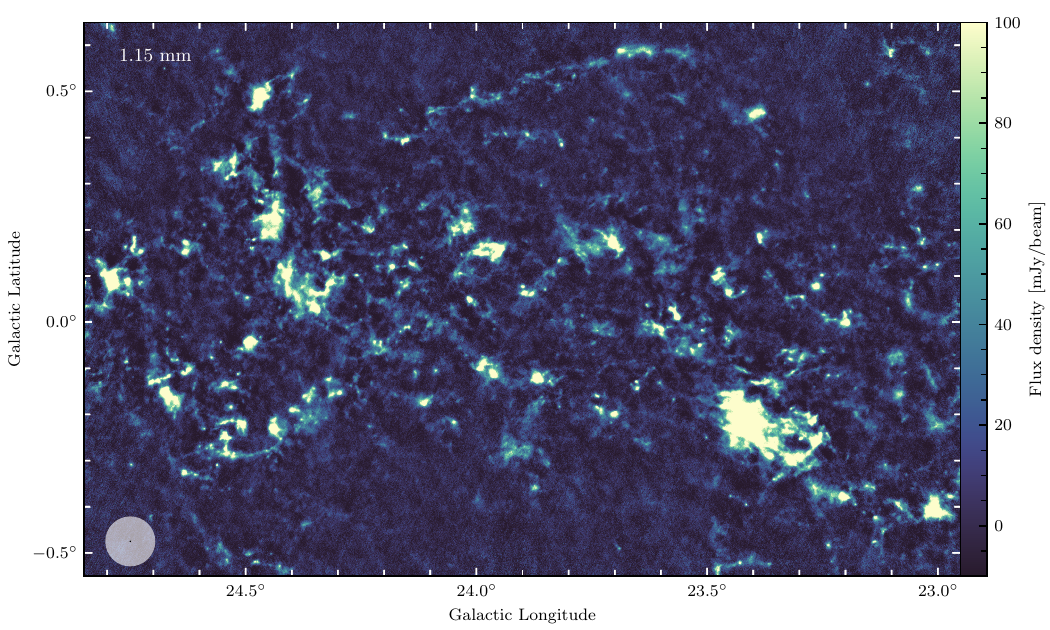}
\label{fig:map1mm}       
\vspace{-7mm}
\caption{The GASTON GP field at 1.15\,mm. The NIKA2 field of view and beam size are indicated by the grey and black circles, respectively, in the lower-left corner.}
\end{center}
\vspace{-7mm}
\end{figure*}

We simultaneously observed a sector of the Galactic Plane in 1.15 and 2.00 mm continuum wavebands using NIKA2 \citep{Bourrion2016, Calvo2016, Adam2018, Perotto2020}, which have angular resolutions of 11.1 and 17.6 arcsec, respectively. The region is centred on $\ell = 23.9^\circ, b=0.05^\circ$, and approximately covers an area of 2 deg$^2$ with fairly uniform sensitivity. The observations consist of pairs of scans centred on three different positions ($\ell = 23.3^\circ, 23.9^\circ, 24.5^\circ$, with $b=0.05^\circ$) with position angles alternating between $\pm 30^\circ$ from the normal to the Galactic plane. Flux calibration is performed via observations of primary calibrators, mainly Uranus and Neptune, and pointing uncertainties are $< 3$ arcsec. The data were reduced using the NIKA2 collaboration's IDL pipeline. The results presented here were obtained from the 1.15 mm image obtained after approximately 30 hours of observations between October 2017 and February 2019. As of August 2021, the GP observations are 92\% complete, out of their total 73.6 hours of allocated time. We display the 1.15 mm map in Figure~\ref{fig:map1mm}. 

\section{Source extraction and determination of physical properties}
\vspace{-1mm}
\label{analysis}

We extracted sources within the 1.15 mm map using the Python {\sc astrodendro} implementation of dendrograms \citep{Rosolowsky2008}, after smoothing the image to 13 arcsec resolution to slightly increase the signal-to-noise ratio. The global rms of the 1.15\,mm map, determined from the null maps produced by the IDL pipeline, is 4.18 mJy per 13-arcsec beam in 2.5-arcsec pixels. We catalogued 488 independent emission structures (dendrogram trunks), containing a total 1467 compact sources (leaves) that contain no further discernible substructures, that will be the focus of our analysis.

We determined systemic velocities for all sources by first creating an average spectrum within each source footprint in overlapping 21 arcsec-resolution $^{13}$CO (1--0) data from the FUGIN survey \citep{Umemoto2017}, and assigned the velocity corresponding to the brightest emission. Secondly, we revised these velocities for sources which also have overlapping coverage in the RAMPS pilot study data \citep{Hogge2018}. These observations of the NH$_3$ (1,1) inversion transition are more appropriate for tracing dense gas, and the observations show better morphological agreement with the 1.15 mm structures than the FUGIN data, although the lower angular resolution (32 arcsec) will probably result in incorrect velocities in some cases. We will eventually revise these velocities with dedicated follow-up observations of dense gas tracers such as N$_2$H$^+$ (2--1). Velocities were then converted to kinematic distances by adopting a Galactic rotation curve model \citep{Reid2019}, and the assumption that the sources are following circular orbits around the Galactic centre. We used version 2.4.1 of the Bayesian distance calculator \citep{Reid2016}, to determine kinematic distance solutions for all of our sources, though we removed the spiral arm model's influence on the resulting distances since we do not know \textit{a priori} that all our sources belong to the spiral arms. We also directly adopted maser parallax distances where possible, in place of the less accurate kinematic distances. Finally, we use a combination of friends-of-friends algorithm, and information contained within the hierarchical structure of the dendrogram, to refine the distances for sources whose velocities were less well defined. In total, we place 92 percent of the compact sources at the near kinematic distance solution, with a median distance of 5.6 kpc. Around 90\% of the sample lies between 3.5 and 7.0 kpc.

\begin{figure}[t]
\vspace{-7mm}
\begin{center}
\sidecaption
\includegraphics[scale=0.8]{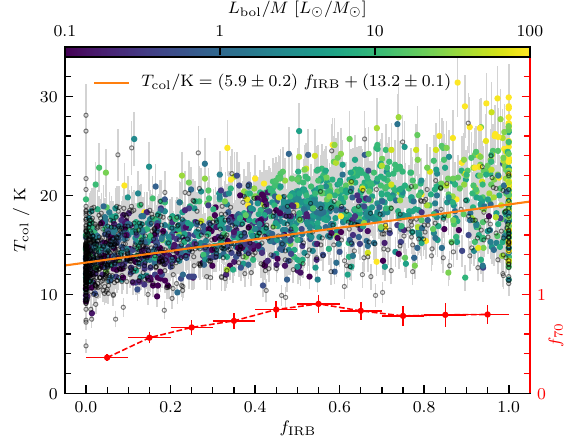}
\label{fig:fIRB}
\caption{The infrared-bright fraction, $f_\mathrm{IRB}$, of all compact sources as a function of 160/250~$\mu$m temperature, coloured by the bolometric luminosity to mass ratio. Sources with no compact 70~$\mu$m counterpart are shown as black circles. On the secondary axis, we show the fraction of sources within 0.1-wide $f_\mathrm{IRB}$ bins that have a counterpart compact 70~$\mu$m source.}
\end{center}
\vspace{-7mm}
\end{figure}

Masses were determined by assuming a modified blackbody for the continuum emission, with dust temperatures estimated using the \textit{Herschel} 160/250 $\mu$m flux density ratio method \citep{Rigby2018}, and a dust absorption coefficient of $0.1 (\nu/\mathrm{THz})^{1.8}$ \citep{Beckwith1990}. The integrated flux densities of each source were first background-subtracted to remove the contribution from overlapping structures in complex regions. We also define a new quantity for each source, the \textit{infrared-bright fraction} ($f_\mathrm{IRB}$), giving the fraction of pixels within the source that are brighter in 8 $\mu$m \textit{Spitzer}/GLIMPSE \citep{Churchwell2009} imaging than the median value within a 4.8-arcmin box (the size of which approximates the largest structures within the GLIMPSE data). We use $f_\mathrm{IRB}$ as a tracer of the relative evolutionary stage of each source, since very young sources tend to be infrared-dark (i.e. $f_\mathrm{IRB} \approx 0$), while sources in an advanced stage of star formation tend to be infrared-bright ($f_\mathrm{IRB} \approx 1$). In Figure \ref{fig:fIRB}, we demonstrate the utility of $f_\mathrm{IRB}$ as an evolutionary tracer, which correlates well with temperature, as well as the bolometric luminosity-to-mass ratio, and likelihood of a coincident compact 70~$\mu$m source \citep{Elia2017}.

\section{Mass and temperature evolution}
\vspace{-1mm}

We refer to the compact sources within our catalogue as \textit{clumps}, though we note that, given the range in deconvolved sizes ($\sim 0.05$--2\,pc) and background-subtracted masses ($\sim 1$--$5 \times 10^3$\,M$_{\odot}$),  our sample suggests that it contains a mixture of both cores (i.e. the precursors of individual star systems) and clumps (i.e. the precursors of stellar clusters). 

\begin{figure*}
\vspace{-7mm}
\begin{center}
\includegraphics[width=\textwidth]{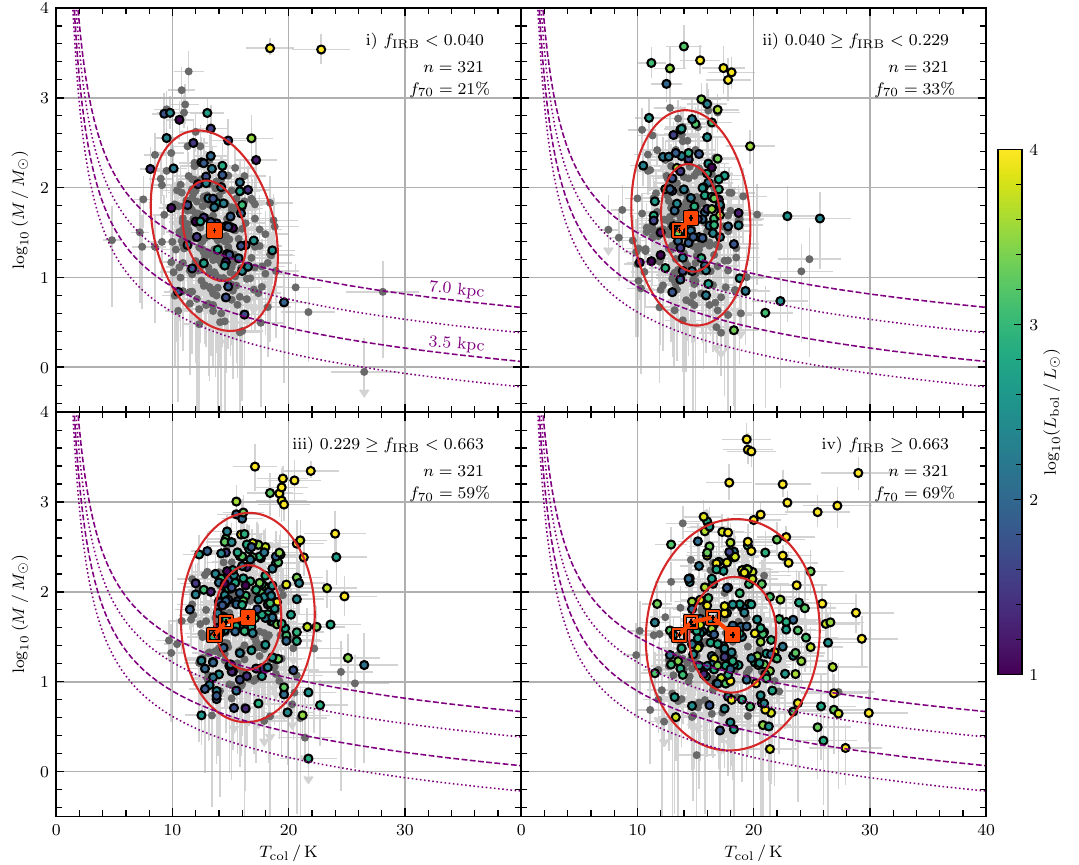}
\label{fig:MTfIRB}       
\vspace{-7mm}
\caption{Clump mass as a function of dust temperature for the four distance-limited subsamples of clumps. For each subsample, the mean value in both axes is shown as an orange square, along with the associated error, and approximate 1- and 2-$\sigma$ confidence ellipses are shown in red. Sources with a compact 70~$\mu$m counterpart are coloured by the corresponding bolometric luminosity, while those without are shown as grey circles. The sensitivity limits for point-like and extended sources are shown as purple dashed and dotted curves, respectively.}
\end{center}
\vspace{-7mm}
\end{figure*}

To quantify the evolution of the properties of the sources, we create a distance-limited sample of the clumps, selecting only those with distances of $3.5 \le d \le 7.0$~kpc, and then break that sample into four equal-sized subsamples (labelled i--iv) depending on their $f_\mathrm{IRB}$ values. In Figure \ref{fig:MTfIRB}, we display the (logarithmic) mass--temperature distributions for four subsamples, and a number of important points are evident. Firstly, the mean temperature increases monotonically as we progress through the samples from infrared-dark to infrared-bright, reflecting the evolution of the underlying star formation. Secondly, we notice that the mean value of the mass is significantly enhanced (when considering the errors on the mean) in stages ii) and iii) compared to i) and iv). Finally, the scatter also increases monotonically.

We compare what we see in Figure \ref{fig:MTfIRB} to models of core- and clump-fed evolution \citep{Peretto2020} in mass--temperature space. For core-fed evolution, we would expect to see sources maintaining a constant mass at early stages in their evolution, before being dispersed by feedback as star formation progresses. By contrast, if the evolution is clump-fed, we expect to see sources gaining mass at early stages, before losing mass at the later stages. The increases of the mean mass in stages (ii) and (iii) compared to (i), which are significant at the 3.1- and 4.3-$\sigma$ levels, respectively, suggests that the populations, as a whole, are gaining mass between these phases, and then losing mass, on average, between stage (iii) and (iv). This would be inconsistent with a population of sources undergoing purely core-fed star-formation, and so we interpret this as evidence that sources are accreting material from their wider environments in a clump-fed scenario.

\section{Summary}
\vspace{-1mm}

We have presented an analysis of deep 1.15\,mm continuum mapping of $\sim$2 deg$^2$ of the Galactic plane centred on $\ell = 24^{\circ}$ using NIKA2. After approximately 40\% of the target integration time, we have identified almost 1500 compact sources after smoothing to 13 arcsec resolution, and identified systemic velocities and distances, and thereby calculated physical properties. By splitting a distance-limited sample into subsamples based on the $f_\mathrm{IRB}$ evolutionary indicator, we find evidence for the mass growth of the compact sources, supporting models of clump-fed star formation in which material is accreted from the wider environment. We expect observations for the GP survey to be concluded during the Winter of 2021--22, enabling the production of maps that are at least twice as deep as those presented in this study, and therefore enhancing the statistical power of this study.

\section*{Acknowledgements} \label{ack}
\vspace{-1mm}
AJR and NP would like to thank the STFC for financial support under the consolidated grant numbers ST/N000706/1 and ST/S00033X/1, and the Royal Society for computing resources under Research Grant number RG150741. We would like to thank the IRAM staff for their support during the campaigns. The NIKA2 dilution cryostat has been designed and built at the Institut N\'eel. In particular, we acknowledge the crucial contribution of the Cryogenics Group, and in particular Gregory Garde, Henri Rodenas, Jean Paul Leggeri, Philippe Camus. This work has been partially funded by the Foundation Nanoscience Grenoble and the LabEx FOCUS ANR-11-LABX-0013. This work is supported by the French National Research Agency under the contracts "MKIDS", "NIKA" and ANR-15-CE31-0017 and in the framework of the "Investissements d’avenir” program (ANR-15-IDEX-02). This work has benefited from the support of the European Research Council Advanced Grant ORISTARS under the European Union's Seventh Framework Programme (Grant Agreement no. 291294). F.R. acknowledges financial supports provided by NASA through SAO Award Number SV2-82023 issued by the Chandra X-Ray Observatory Center, which is operated by the Smithsonian Astrophysical Observatory for and on behalf of NASA under contract NAS8-03060. 

%
%
%

\end{document}